\documentclass[usenatbib,usegraphicx]{mn2e}

\title[Sub-halos, Hot and Cold Gas in a Lyman-$\alpha$ Blob]{Hot Gas, Cold Gas and Sub-Halos in a Lyman-$\alpha$ Blob at Redshift 2.38}
\author[Paul. J. Francis et al.]{Paul. J. Francis$^{1}$\thanks{E-mail:
paul.francis@anu.edu.au}, Michael A. Dopita$^{1,2}$, James W. Colbert$^{3}$, Povilas Palunas$^{4}$,\newauthor Claudia Scarlata$^{5}$, Harry Teplitz$^{3}$, Gerard M. Williger$^{6,7,8}$ and Bruce E. Woodgate$^{9}$\\
$^{1}$Research School of Astronomy and Astrophysics, The Australian National University, Canberra ACT 0200, Australia\\
$^{2}$Astronomy Department, King Abdulaziz University, P.O. Box 80203, Jeddah, Saudi Arabia\\
$^{3}$Spitzer Science Center, California Institute of Technology, Pasadena, CA 91125, USA\\
$^{4}$Las Campanas Observatory, La Serena, Chile\\
$^{5}$Minnesota Institute for Astrophysics, School of Physics and Astronomy, University of Minnesota, Minneapolis, Minnesota 55455, USA.\\
$^{6}$Department of Physics \& Astronomy, University of Louisville, Louisville, KY 40292, USA\\
$^{7}$Lab. Lagrange, U. de Nice, UMR 7293, 06108 Nice Cedex 2, France\\
$^{8}$Inst. for Astrophysics and Computational Sciences, Catholic U. of
  America, Washington DC 20064, USA\\
$^{9}$NASA Goddard Space Flight Center, Greenbelt, MD 20771, USA
}

\begin{document}

\date{Draft 6 Sept 2012}

\pagerange{\pageref{firstpage}--\pageref{lastpage}} \pubyear{2012}

\maketitle

\label{firstpage}

\begin{abstract}
We present integral field spectroscopy of a Lyman-$\alpha$ blob at redshift 2.38, with a spectral resolution three times better than previous published work. As with previous observations, the blob has a chaotic velocity structure, much of which breaks up into multiple components.  Our spectroscopy shows, however, that some of these multiple components are extremely narrow: they have velocity widths of less than $100\ {\rm km\ s}^{-1}$.

Combining these new data with previous observations, we argue that this Lyman-$\alpha$ blob resides in a dark-matter halo of around $10^{13} {\rm M_{\odot}}$. At the centre of this halo are two compact red massive galaxies. They are surrounded by hot gas, probably a super-wind from merger-induced nuclear starbursts. This hot gas has shut down star formation in the non-nuclear region of these galaxies, leading to their red-and-dead colours.

A filament or lump of infalling cold gas is colliding with the hot gas phase and being shocked to high temperatures, while still around 30kpc from the red galaxies. The shock region is self-absorbed in Lyman-$\alpha$ but produces C~IV emission.

Further out still, the cold gas in a number of sub-halos is being lit up, most likely by a combination of tidally triggered star formation, bow-shocks as they plough through the hot halo medium, resonant scattering of Lyman-$\alpha$ from the filament collision, and tidal stripping of gas which enhances the Lyman-$\alpha$ escape fraction. The observed Lyman-$\alpha$ emission from the Blob is dominated by the sum of the emission from these sub-halos.

On statistical grounds, we argue that Lyman-$\alpha$ blobs are not greatly elongated in shape, and that most are not powered by ionisation or scattering from a central active galactic nucleus or starburst.
\end{abstract}


\section{Introduction}

Since their first discovery \citep{Francis:1996p627,Steidel:2000bw}, Lyman-$\alpha$ blobs have held out the prospect of allowing us to study the spatially resolved flows of gas around massive young galaxies. While some Lyman-$\alpha$ blobs surround high-redshift radio galaxies, and are clearly associated with AGN jets \citep[e.g.][]{Reuland:2003gz,Reuland:2007et}, many are not. For these, a wide range of models have been proposed to explain their extended morphology and immense luminosities, including superwinds \citep[e.g.][]{Taniguchi:2001fq}, infalling gas \citep[e.g.][]{Francis:1996p627,Dijkstra:2006dh} and resonant scattering \citep[e.g.][]{Steidel:2011jk}.

Integral field spectroscopy is an ideal way to map the velocity structure in these low-surface-brightness objects. Such spectroscopy has been published for four Lyman-$\alpha$ blobs to date, and all show a similar pattern. LAB1 \citep{Bower:2004p577,Weijmans:2010p583} shows no large-scale systematic velocity field, and multiple velocity components. Many, but not all of the Lyman-$\alpha$ components seen here are seemingly associated with galaxies. 
\citet{Yang:2011hz} observed blobs CDFS-LAB1 and CDFS-LAB2. They too find that the blobs break up into multiple velocity components, which seem to be associated with galaxies. They also find (for CDFS-LAB2) that the Lyman-$\alpha$ emission is self-absorbed at the systemic redshift, with most emission coming at redder wavelengths. \citet{Wilman:2005p582} observed LAB2 and saw absorption at one wavelength crossing their entire field, which they interpreted as absorption in a shell of gas swept up by a superwind, but an alternative interpretation is that it is due to a self-absorbing cloud of gas ionised from within, along the lines suggested by \citet{Zheng:2002p624}. Otherwise LAB2 shows the same multiple emission components seen in LAB1.

Thus all four of the Lyman-$\alpha$ blobs observed to date with spatially resolved spectroscopy show the same pattern: spatially extended but spectrally narrow emission components, often associated with stars, with no clear systematic velocity structure.

In this paper we present integral field spectroscopy of the first Lyman-$\alpha$ blob found, (LAB1$\_$J2143−4423, hereafter B1), originally discovered by \citet{Francis:1996p627}. Our observations have a spectral resolution at the wavelength of Lyman-$\alpha$ four times better than that of the previous observations, which turns out to be unexpectedly informative.

In an appendix, we also present spectroscopy of other Lyman-$\alpha$ blob candidates in the same region of the sky.

Throughout this paper we adopt a cosmology with $H_0 = 71\ {\rm km\ s}^{-1}{\rm Mpc}^{-1}$, $\Omega_{M} = 0.27$ and $\Omega_{\Lambda} = 0.73$, based on the five year WMAP results \citep{Hinshaw:2009jq}. Distances are proper unless otherwise specified.

\section{Target}

The Lyman-$\alpha$ blob B1, originally discovered by \citet{Francis:1996p627} has also been studied in \citet{1997ApJ...482L..25F}, \citet{Francis:2001jb}, 
\cite{Colbert:2006p628} and \citet{Colbert:2p629}. This $z=2.38$ blob remains one of the brightest and best-studied of the Lyman-$\alpha$ blobs. In this section, existing observations of this blob, as reported in the above papers, are summarised.

\begin{figure*}
 \includegraphics*[width=160mm]{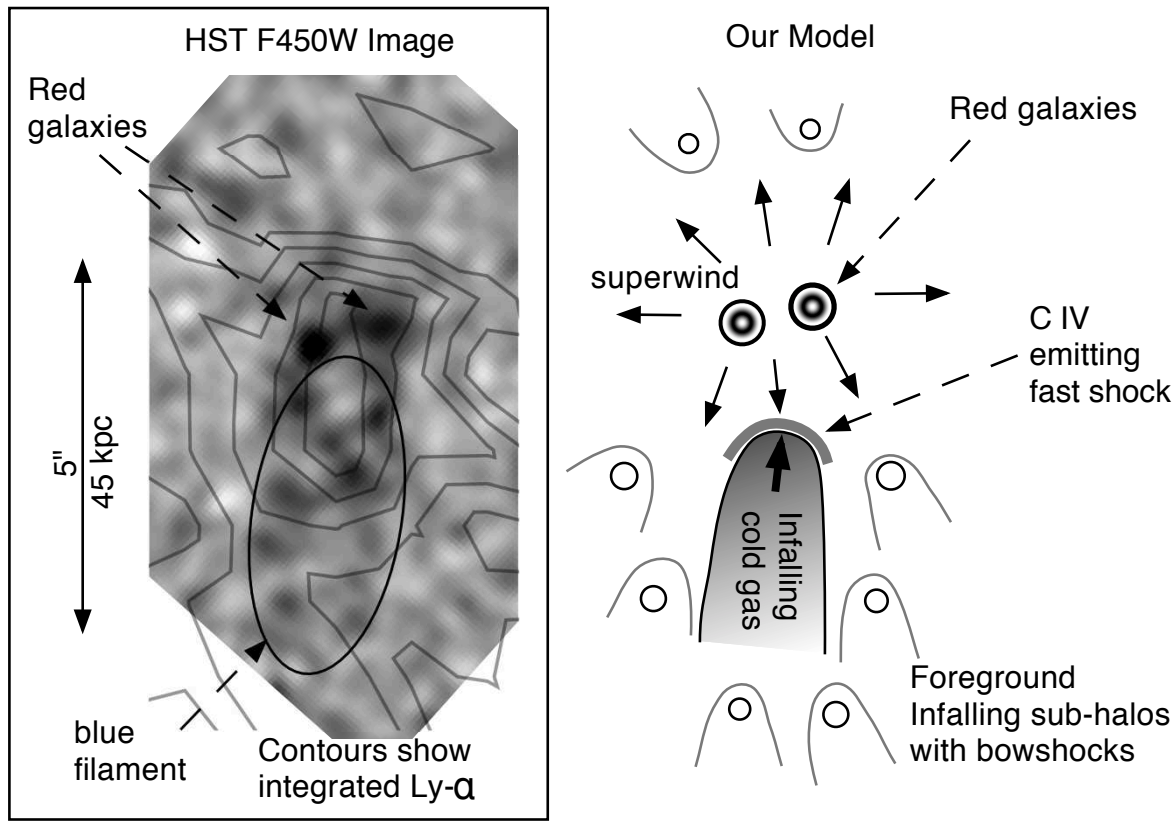}
 \caption{The left panel shows Hubble Space Telescope F450W image of B1 (greyscale) overlaid with the integral Ly$\alpha$ flux measured in this paper (contours, smoothed by a Gaussian 
 with standard deviation $0.5^{\prime \prime}$). Note that previous observations showed that the Ly-$\alpha$ emission continues at low surface brightnesses much further out than shown here. The right panel shows a sketch of our model for B1.
 \label{diagram}}
\end{figure*}

\begin{figure*}
 \includegraphics*[width=160mm]{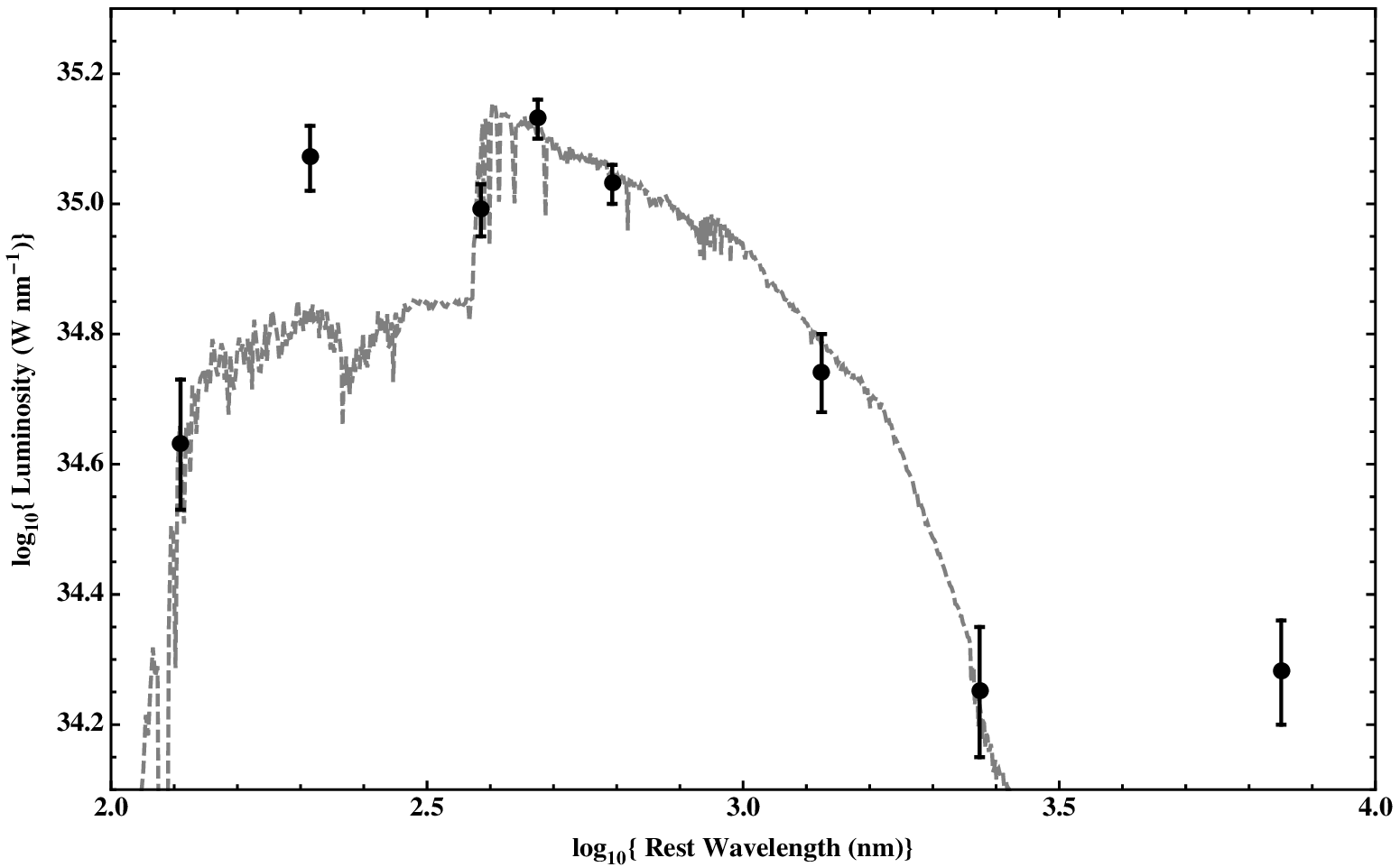}
 \caption{Spectral Energy Distribution of B1 (points) compared with a simple stellar population spectral synthesis model (dashed line) of age $10^7$ years, mass $2.4 \times 10^{11}{\rm M}_{\odot}$ and extinction $E(B-V)=0.37$ mag.\label{SEDplot}}
\end{figure*}

Hubble Space Telescope (HST) imaging \citep{Francis:2001jb} shows that B1 consists of two compact red galaxies separated by $0.8^{\prime \prime}$ (6 kpc), with a faint blue filament extending around 50 kpc to the south (Fig~\ref{diagram}). The blue filament can be explained as a region of ongoing star formation, with little dust and a star-formation rate of around $10 M_{\odot} {\rm yr}^{-1}$  \citep{Francis:2001jb}. The two compact red galaxies are more complicated. \citet{Francis:2001jb} showed that their colours could be explained either by an old ($> 750$ Myr) stellar population, or by a younger one obscured by dust. 

The mid-IR photometry presented by \citet{Colbert:2p629} allows this degeneracy to be resolved (Fig~\ref{SEDplot}). We combined the optimally weighted photometry, summed over all the components, from \citet{Francis:1996p627}, \citet{1997ApJ...482L..25F} and \citet{Francis:2001jb} with the Spitzer photometry from \citet{Colbert:2006p628} and \citet{Colbert:2p629}, to get the spectral energy distribution shown in Fig~\ref{SEDplot}. As the different components of B1 were not resolved in most of these data, the spectral energy distribution is an integrated one. These data were then compared to spectral synthesis models from \citet{Bruzual:2003ck}, combined with dust absorption following the empirical starburst extinction curve of \cite{1994ApJ...429..582C}. The compact red galaxies can be fit by a simple instantaneous burst stellar population of age of 100-500 Myr, dust extinction with $E(B-V) \sim 0.3$ mag, and stellar mass of $ \sim 2 \times 10^{11} M_{\odot}$. They are thus massive but ``red and dead'', with no substantive star formation for at least $10^7$ years. Additional blue emission 
from the filament explains the two shortest wavelength data points. B1 is a much more luminous source  at 24 microns than this model predicts and shows poly-aromatic hydrocarbon (PAH) emission. This implies that a dusty starburst with a star formation rate of around $420 M_{\odot} {\rm yr}^{-1}$ is present, most likely in the core of one or both of the red galaxies \citep{Colbert:2006p628}.

The Lyman-$\alpha$ emission in B1 is genuinely diffuse. Intermediate band HST imaging shows that 13\% of the Lyman-$\alpha$ emission comes from the eastern red compact galaxy, but the rest is not coming from clumps smaller than a few kpc in size, as otherwise we would have seen the clumps.

\begin{figure*}
 \includegraphics*[width=160mm]{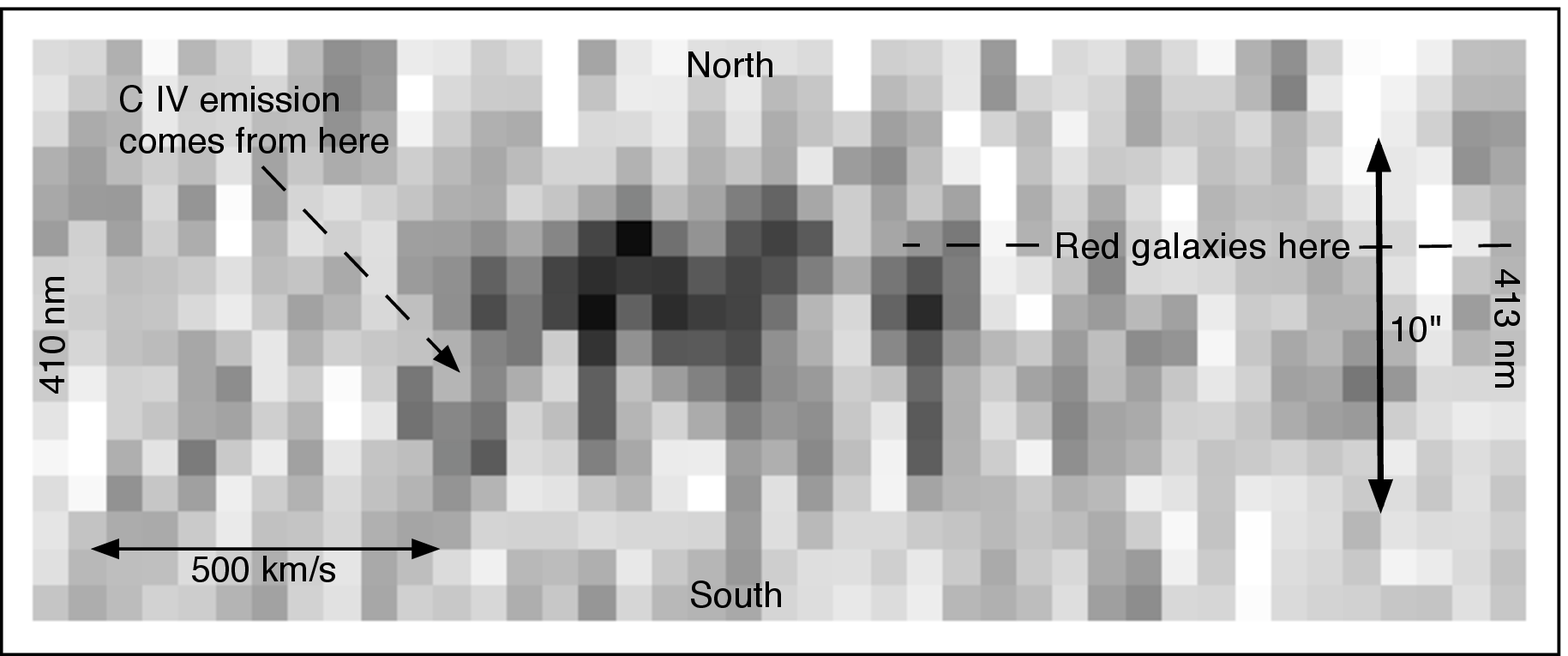}
 \caption{Slice through the data-cube, of width one arcsecond. Wavelength is shown horizontally and position north/south vertically. Spatial pixels are one arcsec wide, spectral pixels are 0.075 nm wide. The data have been spatially smoothed as in Fig~\ref{slicefig}, but not spectrally smoothed.\label{longslitfig}}
\end{figure*}

Strong C~IV (154.9 nm) emission is detected from B1 \citep{Francis:1996p627}. Curiously, it is coming not from the location of the red galaxies, but from 25 kpc south along the filament, and occurs at a wavelength to the blue of most of the Lyman-$\alpha$ emission (Fig~\ref{longslitfig}).

B1 lies at redshift 2.38, in the middle of a supercluster \citep{1993AJ....105.1633F, Palunas:2004p630} traced by QSO absorption lines and Lyman-$\alpha$ emitting galaxies. Spectroscopy of three previously unclassified objects within this supercluster is reported in the Appendix A.

\section{Observations and Reduction}

B1 was observed on the night of August 1 2011, using the WiFeS integral field spectrograph \citep{2007Ap&SS.310..255D, Dopita:2010fc} on the 2.3m telescope at Siding Spring Observatory. WiFeS uses an array of 25 slits, each 38 arcsec long and 1 arcsec wide to obtain integral field spectroscopy of a $25 \times 38$ arcsec field. 
The total integration time on B1 was 4.5 hours. Seeing was around 1.3 arcsec. The B3000 grating was used, giving a spectral resolution of 100 km/s.
The data presented here were reduced using a custom set of scripts written in Mathematica. After bias and overscan subtraction, and flat fielding, the individual slits were traced and straightened. Cosmic rays were identified by comparison with a median of the individual images. Very intense cosmic ray hits (probably the result of beta-decay in the substrate) were found to produce an exponential tail, presumably due to some charge transfer problem. This appeared to be a threshold effect, only occurring for the brightest cosmic rays, and having a constant amplitude and shape for them. These tails were thus easily identified and removed.

Wavelength calibration was done globally, to avoid introducing any row-to-row noise into the reconstructed data cube. Firstly, a polynomial wavelength dispersion relation was found for each spectral row in each slit. A fifth order polynomial was required to fit the arc lines in an individual row. Each of the derived polynomial coefficients was then plotted as a function of slit number and position within each slit. A linear or cubic fit to the variation in each polynomial coefficient was then derived. These fits were then used to reconstruct the wavelength solution for any part of the data. A root-mean-squared residual of 0.013nm was achieved across all arc-lines in all parts of the data, with no systematic patterns to the residuals. Sky lines in reconstructed data cubes were well aligned both along and across slits.

\section{Results}

\begin{figure*}
 \includegraphics*[width=160mm]{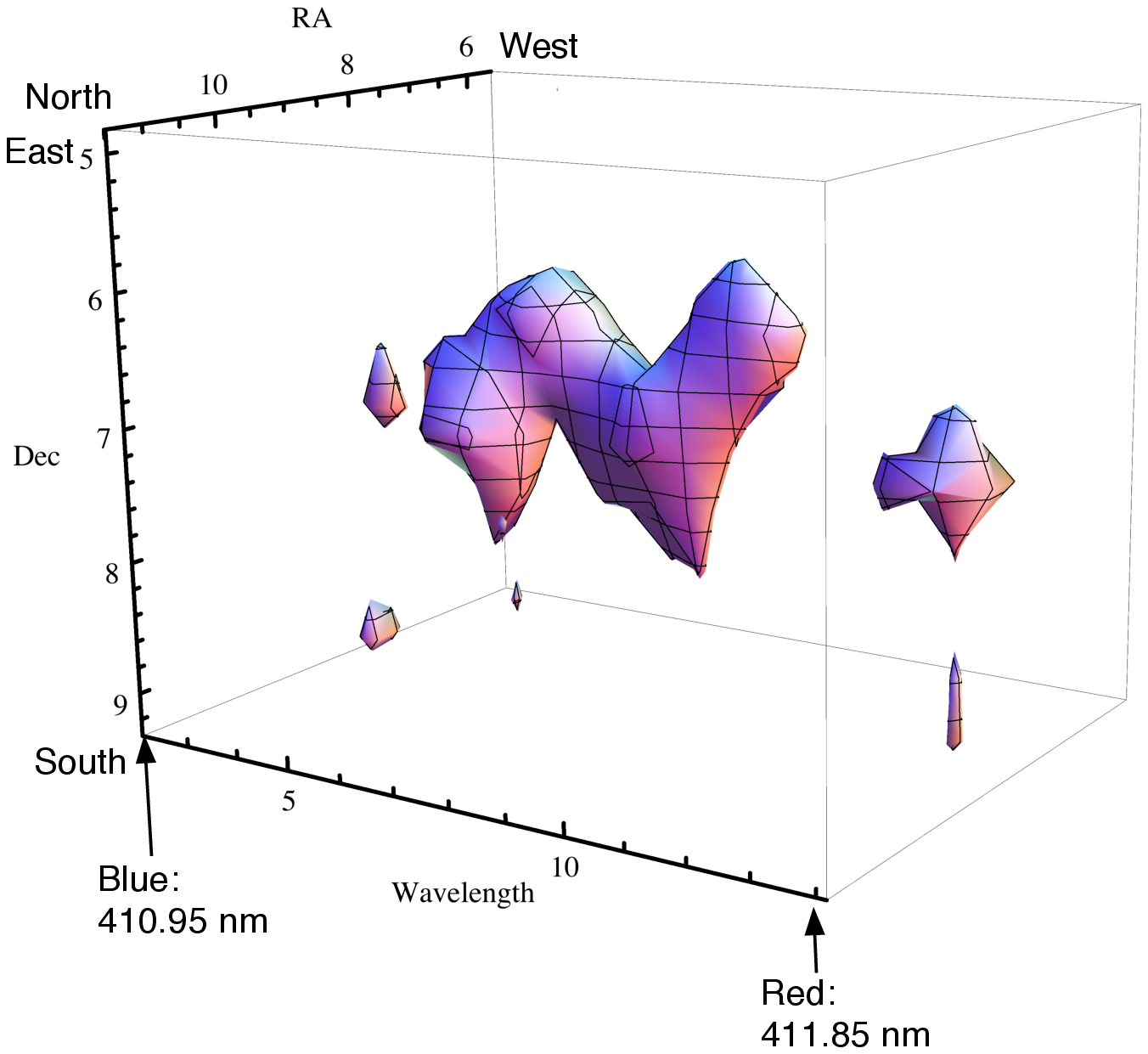}
 \caption{Three dimensional view of the data cube, with the region of strongest flux shown. \label{3dfig}}
\end{figure*}

\begin{figure*}
 \includegraphics*{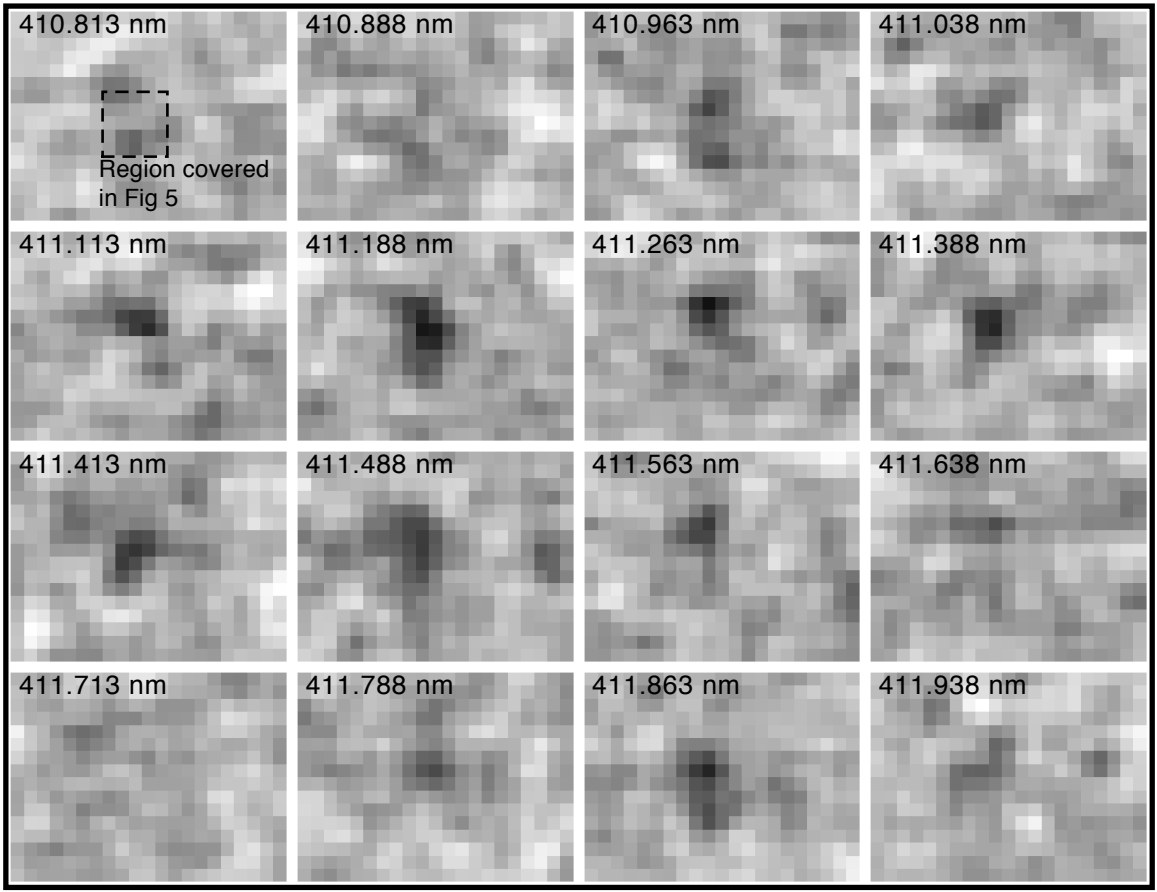}
 \caption{Images of Lyman-alpha emission from B1, as a function of observed-frame wavelength. Each pixel in these images is one arc-second square. North is to the top and East to the left. The images have been smoothed by a Gaussian with a standard deviation of one pixel (one arcsecond). Each slice covers a wavelength range of 0.075 nm. The box in the top-left image shows the field of view of Fig~\ref{specfig}.\label{slicefig}}
\end{figure*}

\begin{figure*}
 \includegraphics*[width=160mm]{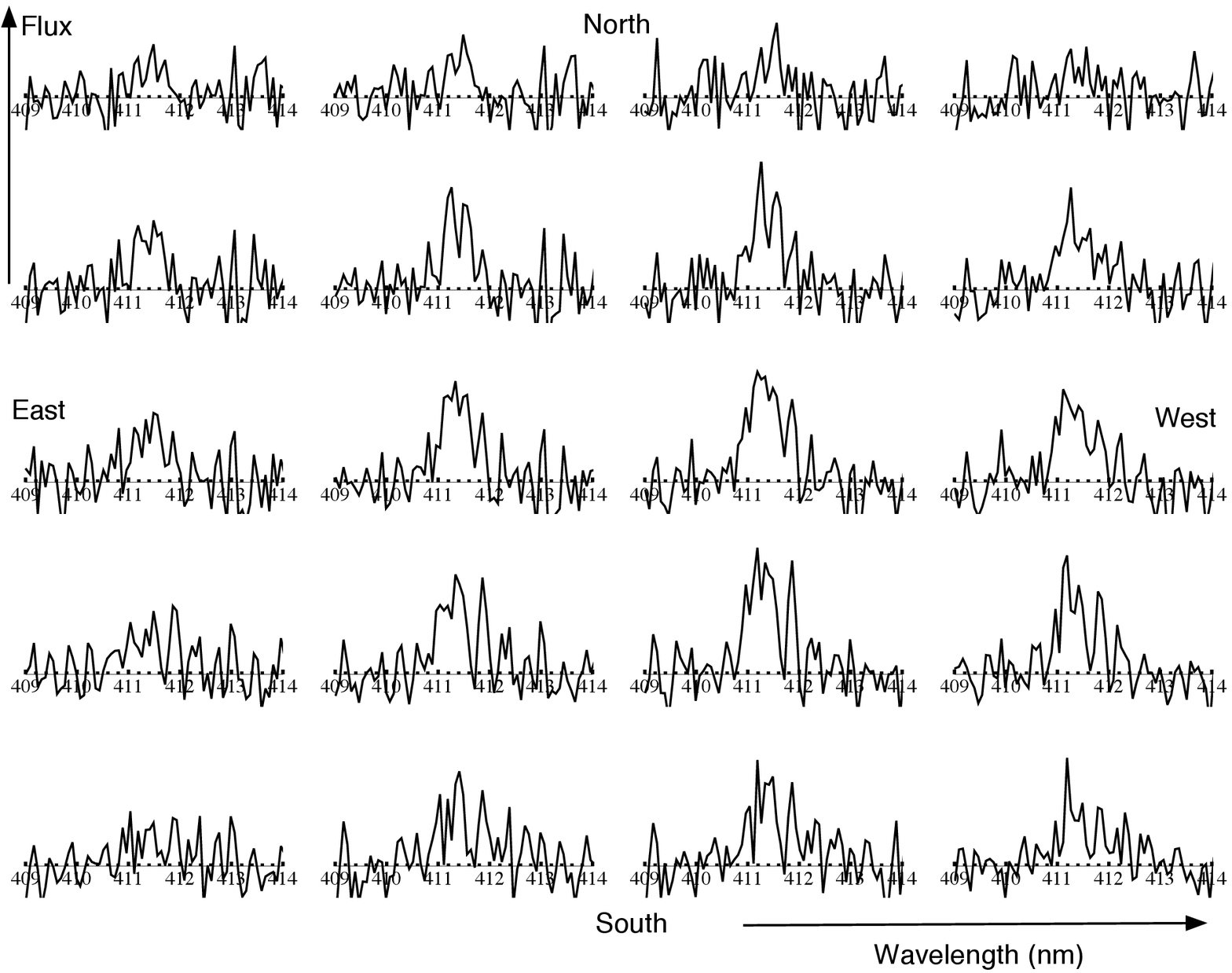}
 \caption{Grid of spectra for individual spatial (1x1 arcsec) pixels. North is to the top and East to the left. The data were spatially smoothed as described in Fig~\ref{slicefig}. The location of this grid is shown in Fig~\ref{slicefig}.\label{specfig}}
\end{figure*}

The only line detected in the WiFeS data was the Lyman-$\alpha$ line of blob B1. Different views of the data cube centred on B1's Lyman-$\alpha$ line are shown in Figs~\ref{3dfig}, \ref{slicefig}, \ref{specfig} and \ref{longslitfig}.

The velocity field is clearly a complex one. A number of conclusions can be drawn:

\begin{enumerate}
\item The Lyman-$\alpha$ emission from B1 is elongated by around 80 kpc in a north-south 
direction. The northern end corresponds to the red galaxies, and the southern end to the blue filament.
\item The peak emission varies as a function of wavelength between the northern and southern ends, giving it a zig-zag appearance in velocity space (Figs~\ref{longslitfig}, \ref{3dfig}).
\item There is a gap in the emission at around 411.7 nm wavelength.
\item The emission, particularly at the southern end, shows a complex velocity structure, with multiple emission and/or absorption components which are spectrally very narrow ($< 100 {\rm km\ s}^{-1}$).
\item The C~IV emission seen by \citet{Francis:1996p627} comes from the southern end of B1,
but at a wavelength blueward of nearly all the Lyman-$\alpha$ emission.
\end{enumerate}

Note that these observations are only picking up the higher surface-brightness parts of B1:
our previous observations showed that low surface-brightness Lyman-$\alpha$ emission extends
far to the north-east of the region discussed here. We also know that a QSO sight-line passing around 160 kpc to the north-west intercepts a Lyman-limit absorption-line system \citep{Francis:2004jg}, indicating that dense neutral hydrogen can be found far beyond the extent of the observable Lyman-$\alpha$ emission.

This zig-zag velocity structure, interrupted by a gap, cannot easily be explained by a 
single rotational, infall or outflow model. It could be explained by a combination of
separate components. Note that it is unlikely that the velocity structures we see are related one-to-one with gas structures. Due to the high optical depth in Lyman-alpha, the emission from any slab of gas structure is likely offset in velocity from its systemic velocity \citep[e.g.][]{Zheng:2002p624,Schaerer:2011p645}, and one slab of gas may produce two or more emission components, typically in the velocity wings. Nonetheless, radiative transfer effects typically broaden lines, so our upper limit on the velocity dispersions of around 100 ${\rm km\ s}^{-1}$ is thus an upper limit on the velocity dispersion in the gas slabs from which this emission arises.

\section{Discussion}

Our observations show a very similar pattern to the previous integral field observations: no simple large-scale systematic velocity field, multiple sub-components, some associated with galaxies. However, our better velocity resolution, combined with the extensive multi-wavelength data-set available for this well studied blob, allows us to draw some somewhat more detailed conclusions than previous authors.

Our discussion starts with a puzzle, an hypothesis and a constraint.

\subsection{Puzzle}

The puzzle concerns gas in B1. The extensive Lyman-$\alpha$ nebula implies the presence of large amounts of cool gas. In contrast, the red-and-dead colours of the compact red galaxies imply that there has been little star formation over a period lasting at least 100 Myr, and hence little cool gas. On the other hand, the mid-IR luminosity and PAH detection indicate prolific star formation and hence large quantities of molecular gas.

The resolution to this puzzle is to have different phases of gas on different scales. The starburst is most likely very compact ($< 1$ kpc) and found in the nuclear regions of one or both of the compact red galaxies. It is so obscured by dust that no sign of it is seen at rest-frame wavelengths shorter than around $2.5 \mu m$ (Fig~\ref{SEDplot}). This nuclear starburst (or starbursts) is probably driven by the interaction between the two red galaxies, which lie at a projected separation of only 6 kpc.
This picture is consistent with the expectations of the \cite{KennicuttJr:1998id} star formation law in which the surface rate of star formation is related to the surface density of gas though $\Sigma_{\rm SFR} \propto \Sigma_{\rm gas}^{1.5}$. Regions of high specific star formation are therefore regions of high gas (and dust) surface densities, and are therefore likely to be heavily dust-obscured. The Lyman-$\alpha$ emission we detect from one of the red galaxies would presumably be  coming from the outskirts of the galaxy.

On scales of a few kpc, the gas is probably too hot to allow star formation. This explains the red-and-dead colours of the compact red galaxies. A superwind is likely present, as nuclear starbursts as luminous as that in B1 virtually always produce superwinds in the local universe \citep[e.g.][]{Veilleux:2005ec}. This could have swept out any cool gas and truncated star formation in the outer regions of these galaxies. \citet{Taniguchi:2001fq} suggested such superwinds as power-sources for Lyman-$\alpha$ blobs. The age of the stellar population ($\sim 10^8$ years) 
is comparable to the merger timescale, and the current star-formation rate, acting over 
$10^8$ years would produce a stellar mass comparable to that observed. The gas may also,
however, have been heated by a concealed active galactic nucleus \citep[e.g.][]{King:2011ey} 
or by the liberation of 
potential energy from infalling gas \citep[e.g.][]{Johansson:2009bb}.

We now turn to the neutral hydrogen responsible for the Lyman-$\alpha$ nebula. This presumably lies further out, and is the subject of our hypothesis. 

\subsection{Hypothesis\label{simulations}}

Our data show that the extended Lyman-$\alpha$ flux is typically tens of kpc from the red galaxies (which we assume lie at the centre of the dark matter halo), spatially extended on scales of several kpc, but breaks up spectrally into narrow sub-components. All these properties sound similar to those of a family of dwarf galaxies surrounding a massive galaxy. Which led us to the following hypothesis: that much of the extended Lyman-$\alpha$ flux is coming from neutral gas within dark-matter sub-halos, moving within the much more massive dark matter halo centred on the red galaxies.

How feasible is this model? The relative velocities of the sub-clumps are $\sim 700 {\rm km\ s}^{-1}$, and they are spread over $\sim 50 {\rm kpc}$, which gives a virial mass of the main halo of $\sim 10^{13} M_{\odot}$. We selected ten such massive halos from the Millenium II n-body simulation \citep{BoylanKolchin:2009co} at this redshift, and looked at the sub-halo population of these ten big halos. There were typically 10 -- 20 sub-halos within each big halo. They were spread over around 100 kpc, and had a line-of-sight velocity dispersion of around $500 {\rm km\ s}^{-1}$. And the velocity dispersions of the individual sub-halos were typically around $70 {\rm km\ s}^{-1}$. Thus, the observational and the predicted parameters appear to match each other very well. These sub-halos in the simulation had typical dark matter masses of $\sim 10^{11} M_{\odot}$ or more and half-mass radii of 1 -- 5 kpc.

Would one expect sub-halos such as these to contain neutral hydrogen? We picked out a sample of small halos with masses of $\sim 10^{11} M_{\odot}$ from the semi-analytic models of \citet{DeLucia:2007ju} applied to the Millenium I simulation \citep{Springel:2005gv}, and they typically had cold gas masses of $\sim 10^9 M_{\odot}$. So current semi-analytic models do suggest that at these redshifts, halos with masses of $\sim 10^{11} M_{\odot}$ will contain reasonable amounts of neutral gas.

Why are these sub-halos radiating Lyman-$\alpha$ emission? The sub-clumps we observe have Lyman-$\alpha$ luminosities of $\sim 2 \times 10^{35} W$, which are well within the range 
found in observed Lyman-$\alpha$ emitting galaxies. However, the space density of $\sim 10^{11} M_{\odot}$ halos is vastly greater than the space density of Lyman-$\alpha$ emitting galaxies \citep[e.g.][]{Guaita:2010kr}. Thus most such halos do not radiate at this level. The equivalent widths are also different: we compared the equivalent widths of blobs with those of compact Lyman-$\alpha$ emitting galaxies in two surveys: \citet{Palunas:2004p630} and \citet{Matsuda:2004p631}. Both surveys measured the equivalent widths (or equivalently narrow-band minus broad-band colour) for both the blobs and other Lyman-$\alpha$ emitting galaxies in their fields, and in both cases, blobs were picked purely on the basis of size in the narrow-band image. In both cases, Lyman-$\alpha$ blobs tend to have higher equivalent widths. A Kolmogorov-Smirnov two-sample test shows that the difference is significant at the 95\% confidence level for the Palunas data, and at the 99.7\% confidence level for the larger Matsuda dataset (using objects brighter than a narrow-band magnitude of 24.2). We therefore conclude that Lyman-$\alpha$ blobs have significantly higher equivalent widths.

Thus for the hypothesis to be valid, we need some mechanism which increases the Lyman-$\alpha$ luminosity and equivalent width of $\sim 10^{11} M_{\odot}$ sub-halos when they are within the dense environment of massive halo, compared to the values seen in similar mass halos in less dense environments. We will discuss the possibilities in Section~\ref{bright}.

Note that some of these sub-halos may contain stars at this redshift. Sensitive observations may thus find that most, but probably not all of the Lyman-$\alpha$ clumps have faint galaxies within them. \citet{Prescott:2011wl} recently presented imaging showing that one Lyman-$\alpha$ blob does indeed seem to comprise a compact group of dwarf galaxies, though in this case the galaxies were not clearly associated with the Lyman-$\alpha$ emission. \citet{Weijmans:2010p583}, however, did find that most clumps of Lyman-$\alpha$ emission were centred on faint galaxies. So this is plausible.

\subsection{The C~IV Constraint}

We now come to our constraint: the strong C~IV (154.9 nm) emission detected coming from $\sim 25$ kpc south of the red galaxies. As discussed in \citet{Francis:1996p627}, C~IV requires a hard ionising spectrum and must therefore either come from Active Galactic Nucleus (AGN) photoionisation or from fast shocks. We will discuss these possibilities in turn.

\subsubsection{AGN Model}

For the AGN model to work, we would require that at least one of the red galaxies hosts a type-2 AGN. Our line-of-sight to the AGN is blocked by a dusty torus, but UV radiation is escaping at least in the southerly direction, and impacting on some relatively dense gas cloud, where it produces Lyman-$\alpha$ and C~IV via photoionisation \citep[e.g.][]{Haiman:2001bb,Adelberger:2006ii,Francis:2006cya}. A second mechanism,  excitation by collision with a radio jet, is ruled out by our upper limit on the radio flux from B1 \citep{Francis:1996p627}. 

The first argument against this model is the detection of PAH emission from B1, which suggests that if a concealed AGN is present, its energy output is dominated by that from star formation. Note however that the PAH emission was only tentatively detected \citep{Colbert:2p629}, and that even an AGN which makes no significant contribution to the total energy output could still produce
significant photoionisation.

A second argument against the AGN model is statistical. 
If the AGN model is correct, and the ratio of obscured to unobscured AGN is $r$, then if we see $n$ blobs in a field, we should see $n/r$ unobscured AGN. We know that r is not significantly greater than 3 \citep[e.g.][]{Webster:1995fb, Francis:2004ev,Treister:2009p653}.

In the Palunas et al. field, we have three (or four if you count B5) Lyman-$\alpha$ blobs. How many QSOs lie in this field at our redshift?
Two QSOs were identified at the cluster redshift, one with $B=21.24$ at coordinates 21:42:31.33 -44:30:16.8 found by \citet{Francis:2004ev}, and B29 identified by our GMOS observations described in Appendix~A, with $B=23.16$. In the field surveyed by 
\citet{Matsuda:2004p631}, there are 35 Lyman-$\alpha$ blobs, and five known QSOs with optical magnitudes in the range 20.5 to 24.5, found in the NASA  Extragalactic Database. In the survey by 
\citet{Yang:2009p635}, there are four Lyman-$\alpha$ blobs and no QSOs down to the Sloan Digital Sky Survey limit of $g<19.4$.

Combining these samples, we find that we can get suitable numbers of unobscured QSOs to be consistent with the blob population, but only if we include faint QSOs: those down to $V \sim 23$ or fainter. 

Could QSOs this faint feasibly enhance the Lyman-$\alpha$ emission from nearby gas clouds to give the luminosity we observe in Lyman-$\alpha$ blobs? We can answer this empirically by looking at the observed ratio of QSO magnitude to the flux of extended Lyman-$\alpha$ emission around it.
\citet{Weidinger:2005p636} and \citet{Christensen:2006p637} between them detected extended Lyman-$\alpha$ emission around 9 QSOs. The average extended Lyman-$\alpha$ flux was $3.2\times10^{-16}{\rm erg\ cm}^{-2}{\rm s}^{-1}$, which is comparable to Lyman-$\alpha$ blob fluxes, so QSOs certainly can produce extended emission as seen in Blobs. However, the QSOs were very bright: averaging $V=18.2$: much brighter than the QSOs found in the various Blob fields. And many QSOs this bright in the Weidinger and Christensen samples had no detectable extended Lyman-$\alpha$ flux around them, so this average ratio of extended flux to nuclear flux is really only an upper limit. If we assume that the extended flux scales with the nuclear flux, then the QSOs in the blob fields are around two orders of magnitude less luminous than would be needed to produced the observed Lyman-$\alpha$ blob flux. Would we expect the extended flux to be proportional to the 
photoionising luminosity? It is possible that the most luminous QSOs actually suppress the Lyman-$\alpha$ emission of nearby galaxies, as suggested by 
\citet{Francis:2004dqa}. In this case, less luminous QSOs might be surrounded by stronger Lyman-$\alpha$ fuzz than the most luminous ones (which are the only ones studied to date), and hence might be plausible unobscured counterparts of Lyman-$\alpha$ blobs. We note, however, that we did not detect extended Lyman-$\alpha$ emission from either of the QSOs lying within our supercluster.

We therefore conclude that if blobs contain concealed QSOs, and these concealed QSOs are of comparable luminosity to the other QSOs seen in the blob fields, then they are unlikely to be
luminous enough to produce the extended Lyman-$\alpha$ emission through photoionisation. If
the concealed QSOs are much more luminous, then it becomes difficult to understand why
we don't see at least a few comparably luminous unobscured QSOs in these fields, unless 
the radiation from these QSOs is completely blocked, in which case it cannot escape to
photoionise the extended Lyman-$\alpha$ emission. Thus while the AGN model cannot be
ruled out for individual blobs, for statistical reasons it is unlikely that AGN photoionisation
is the dominant power source of the population of Lyman-$\alpha$ blobs.

\subsubsection{Fast Shocks Model}

\citet{Francis:1996p627} and \citet{Francis:2001jb} originally suggested that fast shocks could cause the C~IV emission from B1, and demonstrated that they could produce ample line emission provided the density of the medium into which they were passing was at least $10^{-2} {\rm cm}^{-3}$ and the shocks had become radiative, which required a timescale of $\sim 10^8$ years, which is comparable to the infall time for B1 and hence plausible.

Fast shocks are almost inevitable given the high velocity dispersion and observed lumpiness of the gas around B1. A plausible location for the shocks would be a collision between a superwind from the red galaxies and gas infalling along a filament from the south. 

What is the infalling gas which is being shocked? One possibility is gas within an infalling sub-halo, perhaps an unusually massive one. Another is cold-mode infall
\citep[e.g.][]{FaucherGiguere:2010p608,Goerdt:2010p611}. The infalling gas cannot, however, be primordial, as we need it to contain carbon. The presence of a superwind is not necessary for this model to work: the impact of an infalling cold gas cloud on the hot central medium will suffice to drive suitable shocks.

\subsection{Making the Sub-Halos Glow\label{bright}}

Our model, based on the puzzle, hypothesis and constraint discussed above, is summarised in Fig~\ref{diagram}. To make it work, however, we have to find some mechanism to make the gas in each sub-halo emit a Lyman-$\alpha$ luminosity of $>2 \times 10^{35} W$. There are several plausible mechanisms and we will discuss them in turn.

\subsubsection{Induced Star Formation}

The ram pressure of their passage through the hot medium, or tidal interactions, could perhaps trigger star formation in these sub-halos, which would in turn generate the Lyman-$\alpha$ emission. If we take the observed Lyman-$\alpha$ luminosities, assume a typical unobscured ratio of Lyman-$\alpha$/H$\alpha$ of $\sim 8:1$, and apply the relation between H$\alpha$ luminosity and star formation rate from \citet{KennicuttJr:1998ki}, then we infer star formation rates in each sub-clump of around 2-3 solar masses per year, and hence multiplying by the number of sub-clumps, a total star formation rate of around 10--20 solar masses per year. This would just about be consistent with producing the observed UV emission from the blue filament (which gave an inferred star formation rate of around $10 M_{\odot} {\rm yr}^{-1}$). In practice, however, the optical depth in Lyman-$\alpha$ is likely to be so high that ratios of Lyman-$\alpha$/H$\alpha$ will be closer to  $\sim 1:1$, which would raise the inferred star formation rate to levels that are inconsistent with the faintness of the observed blue continuum flux. Another way to say this is that the equivalent width of the Lyman-$\alpha$ emission is higher than that observed for other Lyman-$\alpha$ emitting galaxies, as discussed in Section~\ref{simulations}. Star formation is thus unlikely to produce all 
the Lyman-$\alpha$ emission, though it could produce some of it.

\subsubsection{Infall-powered emission}

A typical sub-halo, according to semi-analytic models, will have a cold gas mass of $> 10^9 M_{\odot}$, and an infall time of $\sim 10^8$ years. If we divide the gravitational potential energy of the gas in a sub-halo by the infall time, we get a power of $\sim 10^{37} W$, which is 500 times greater than the observed Lyman-$\alpha$ luminosity. Thus if we can convert even 1\% of the gravitational potential energy into radiation, we can explain the brightness of the sub-halos.

One mechanism for doing this \citep[e.g.][]{Dijkstra:2009p602} is cooling radiation of thermally heated gas. Another is shock
ionisation. If we assume that each sub-halo has a neutral gas surface area of several tens of square kpc, then a Lyman-$\alpha$ luminosity density of a few $\times 10^{-6}{\rm W\ m}^{-2}$ is needed to reproduce the observed luminosities. If a shock-wave is driven into this surface as the sub-halo moves through the ambient gas, this shock will produce a diffuse Lyman-$\alpha$ emission.
The model shock grids of \citet{Allen:2008gj} show that such a Lyman-$\alpha$ luminosity density can be easily produced, provided pre-shock densities are $> 0.1 {\rm cm}^{-3}$ and shock velocities are $> 200 {\rm km\ s}^{-1}$. Higher velocities or densities give much higher luminosities, allowing smaller gas
surface areas, or more self-absorption.

Thus infall is a viable model for producing at least a substantial part of the observed Lyman-$\alpha$ luminosity.

\subsubsection{Scattered Lyman-$\alpha$ from a central source}

One possibility is that the emission we see from the sub-halos is actually produced elsewhere, and is just being resonantly scattered by the sub-halos. The emission could be coming from the same region that generates the C~IV emission (according to our model, a region ionised by fast shocks where infalling gas hits a superwind), or escaping from the nuclear starbursts in the red galaxies. We do not see Lyman-$\alpha$ emission directly from these regions, due to their high optical depths, but instead we see part of it scattered into the wings of the velocity distribution and off nearby gas clouds. This mechanism was invoked by \citet{Steidel:2011jk} to explain the faint diffuse Ly$\alpha$ halos seen in stacked images of high redshift galaxies, and can lead to polarised Lyman-$\alpha$ emission \citep[e.g.][]{Dijkstra:2008kd}. \citet{Hayes:2011p606} found that the Lyman-$\alpha$ emission in LAB1 was indeed polarised, though \citet{Prescott:2011p607} did not detect such polarisation in much shallower observations of a different Lyman-$\alpha$ blob.

How plausible is such a model in the case of B1? \citet{Steidel:2011jk} found that the extended Lyman-$\alpha$ flux typically exceeded the compact central line flux by a factor of $\sim 5$, and in our case, the extended Lyman-$\alpha$ flux exceeds the central Lyman-$\alpha$ flux (from one of the red galaxies) by a factor of 7, which is comparable.

If scattered Lyman-$\alpha$ emission originally came from the superwind/infalling gas collision, then the UV emission producing it came from fast shocks (as we know from our C~IV detection) and would hence have a very hard spectrum. This means that we would expect very high equivalent widths and little detectable UV continuum, as observed.

If, on the other hand, the scattered Lyman-$\alpha$ emission originally came from star formation, as proposed by \citet{Steidel:2011jk}, we might expect to see UV emission from this star formation. We can estimate how bright it should be by scaling from the results of \citet{Steidel:2011jk}. B1 has a Lyman-$\alpha$ surface brightness in its central few arcsec which is $\sim 30$ times that of the halos seen by Steidel et al., so one would expect blue continuum emission $\sim 3.5$ magnitudes greater than the $V_{AB} \sim 25$ of the galaxies studied by Steidel - i.e. we should expect $V \sim 21.5$. In fact, B1 has $V \sim 23.5$. The red galaxies in B1 show no evidence in their optical colours of any ongoing star formation: we only see evidence for it in the mid-IR. A similar problem occurs in LAB1, where the probable central galaxy is also red 
\citep{Chapman:2004fq}. In both B1 and LAB1 there is plenty of star formation taking place, which is seen in the mid-IR and sub-mm, but little or no detectable UV light.

To put this another way, the star formation rate inferred from the observed blue light in B1 (the blue filament) of $\sim 10 M_{\odot}{\rm yr}^{-1}$ is insufficient to produce the phenomenal observed Lyman-$\alpha$ luminosity. We would need to tap into the much greater star formation rate which is concealed by dust, most likely in the red galaxies ($\sim 420 M_{\odot}{\rm yr}^{-1}$). 

The only way out of this would be to make the radiation from starbursts anisotropic. We would not see it from our orientation, but it would escape in other directions, analogous to unified models of AGN. If this was the case, we would sometimes expect to see blobs in which the central starburst was exposed. These would look like compact Lyman-$\alpha$ emitters surrounded by relatively faint fuzz, rather than diffuse blobs, but should have Lyman-$\alpha$ luminosities greater than that of blobs. There are two potential exposed starbursts in the Palunas et al. supercluster: B5 (Appendix~A) and B40 \citep{Francis:2004p634}. Both have Lyman-$\alpha$ luminosities and line widths comparable to the three blobs in this field. B5 has a compact nucleus surrounded by relatively faint fuzz, while B40 is spatially unresolved. 
But the extended flux of these objects is much fainter than that of the Lyman-$\alpha$ blobs, so if viewed from some other orientation they might well be blobs, but much fainter ones. There are thus no possible counterparts of the most luminous blobs in this field.

Thus the scattering model is feasible, but only if we are scattering emission generated by fast shocks, rather than by star formation.

\subsection{Tidal streams}

Regardless of the mechanism or mechanisms producing the Lyman-$\alpha$ emission from the sub-clumps, one would expect some tidal stripping of gas from them in this dense environment, perhaps producing Magellenic-stream style gas streams. This extended narrow gas morphology will increase the escape fraction of Lyman-$\alpha$ and help enhance the observed brightness of these sub-halos regardless of the cause of the emission.

\subsection{Intrinsic shape of the Sub-halo Swarm}

If Lyman-$\alpha$ blobs are collections of sub-halos, are these collections roughly spherical or are they elongated? There is some evidence for blobs lying in filamentary structures \citep{Erb:2011p612} but is this reflected in their shapes?

In this section, we develop a simple toy-model of elongated Lyman-$\alpha$ blobs, taking into account a selection effect which biases samples towards finding end-on blobs. We compare this to data on the apparent shapes of Lyman-$\alpha$ blobs from \citet{Matsuda:2004p631}, and show that blobs are not highly elongated on average.

\begin{figure*}
 \includegraphics*{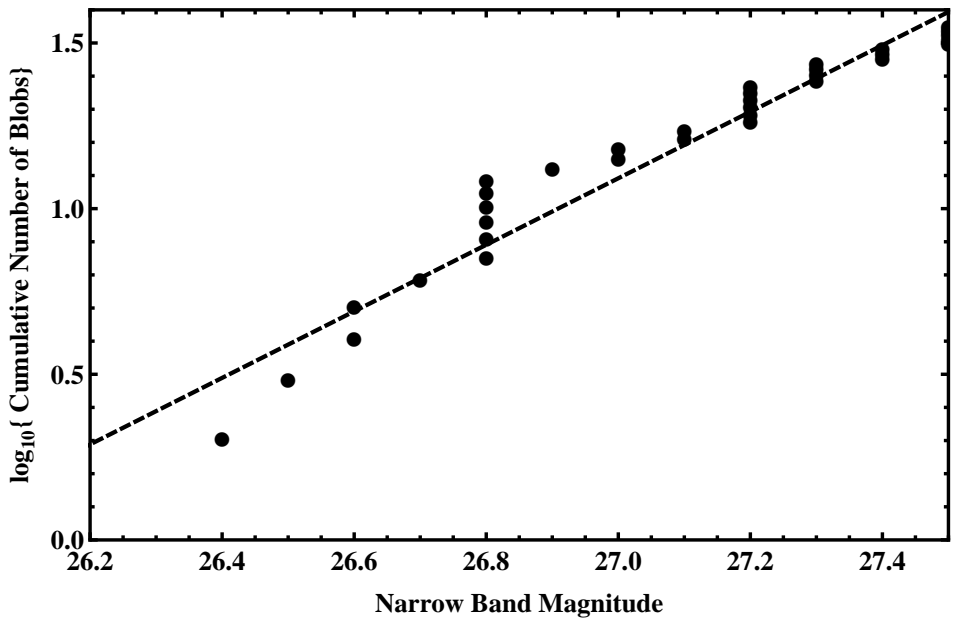}
 \caption{The cumulative number of Blobs found by Matsuda et al, as a function of
 narrow-band magnitude. Each point represents one blob. The line is a best fit model.\label{matsudacounts}}
\end{figure*}

Let us model Lyman-$\alpha$ blobs as cigar-shaped ellipsoidal distributions of line-emitting clouds. To keep the model simple, we assume that the observed Lyman-$\alpha$ surface brightness of a blob is proportional to the length of our line of sight through the ellipsoidal distribution. We further assume that the long axis of the blob is a factor $r$ longer than the minor axis, and that we are observing at an angle $\theta$ to the long axis. Simple geometry then tells us that the observed surface brightness b is given by:
\begin{equation}
b \propto \sqrt{1+(r^2-1) \cos^2(\theta}).
\end{equation}
Any narrow-band search for Lyman-$\alpha$ blobs will have a surface brightness cut-off. Blobs that would normally lie below this cut-off will make it into the sample if viewed end-on, so any sample will be biased towards blobs seen at small values of $\theta$. How big is this bias? To estimate this, we need to 
know how the space density of Lyman-$\alpha$ blobs increases as the surface brightness decreases. To parameterise this, we plotted the cumulative number of blobs found by \citet{Matsuda:2004p631} against their surface brightness, down to the rough completeness limit of this sample (Fig~\ref{matsudacounts}). We fit the cumulative space density in this survey with a straight-line model: the best fit is:
\begin{equation}
\log_{10}{\rm (Cumulative\ Number\ of\ Blobs)} \propto 1.0\ {\rm mag}_{nb}
\end{equation}
Thus if blobs seen end-on are a magnitude brighter than those seen
side-on, the end-on ones will be over-represented in a sample by a factor of around 10.

\begin{figure*}
 \includegraphics*{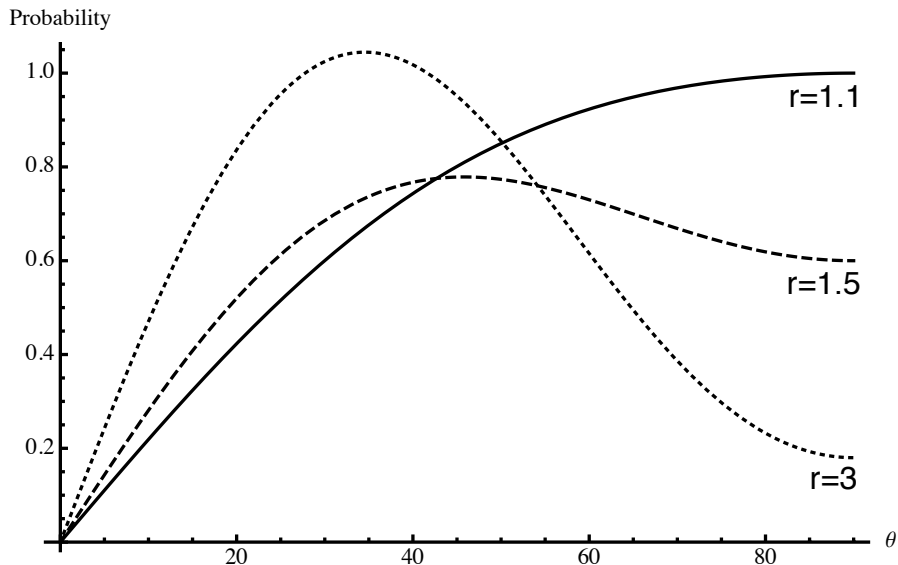}
 \caption{Probability of observing a Lyman-$\alpha$ blob as a function of orientation angle $\theta$, for three different elongations $r$.\label{selectprob}}
\end{figure*}

We assume that the cumulative number density has a power-law dependence on limiting magnitude,
with an index chosen to reproduce this factor of ten. For any given surface-brightness enhancement, we can then calculate how many more blobs we should see. We further assume that in the absence of any selection effects, blobs are
randomly orientated in three dimensions, which means that the probability of seeing a 
blob from a given angle $\theta$ is proportional to $\sin(\theta)$. The observed distribution is then the product of the random orientation with the bias towards end-on blobs, which comes out as:
\begin{equation}
\sin{\theta }\times10^{(1.0 \times 2.5 \log10{\sqrt{1+(r^2-1)\cos^2{\theta}})}}
\end{equation}

which simplifies to
\begin{equation}
\left(1+\left(r^2 - 1 \right) \cos^2{\theta }\right)^{1.25} \sin{\theta }
\end{equation}
which is plotted in Fig~\ref{selectprob}.

Thus even quite modest elongations will have a strong impact on the viewing angle at which Lyman-$\alpha$ blobs are seen. We thus confirm that there can be a bias towards small viewing angles $\theta$.

\begin{figure*}
 \includegraphics*{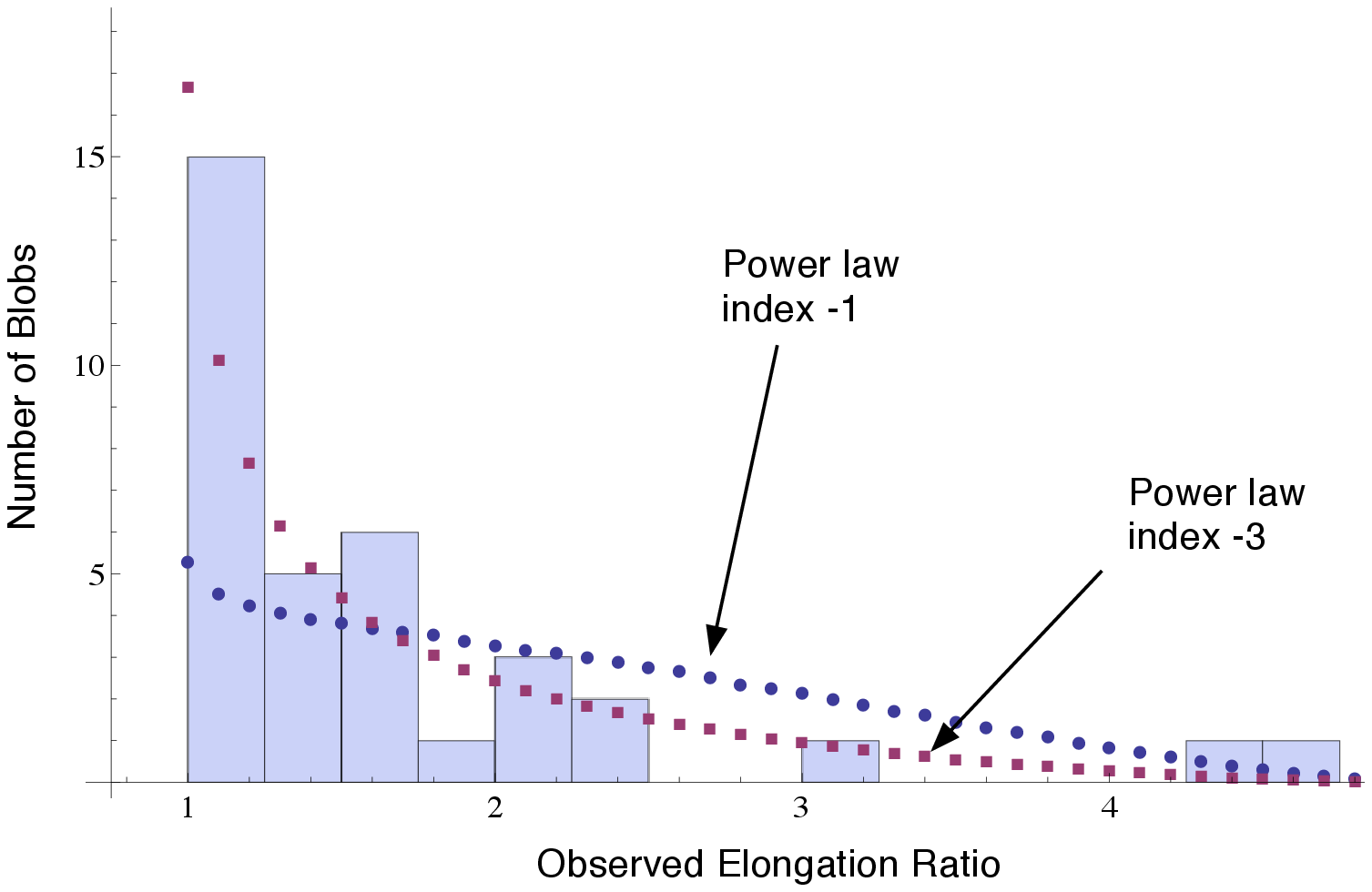}
 \caption{Comparison between the observed (histogram) and modelled (points) distributions of 
 observed axial ratios. The two models are for populations of blobs with power-law distributions of r, with indices -1 and -3.\label{elongations}}
\end{figure*}

If we assume a power-law intrinsic distribution of elongations ($r$ values), we can then do a monte-carlo calculation of 
the distribution of observed axial ratios projected onto the sky (the ratio of the observed long- to short-axis of the blob), and compare this with the observed distribution taken from \citet{Matsuda:2004p631}. The results are shown in Fig~\ref{elongations}.

This figure shows that most of the Lyman-$\alpha$ blobs observed by Matsuda were not greatly elongated. 
The best fit is where the number of blobs with a given intrinsic elongation ($R$) is proportional
to $r^{-3}$. Models in which elongated blobs are more common over-predict the number of
blobs with large observed elongations.

The Millenium II simulations analysed in section~\ref{simulations} are also not strongly elongated. Some of the massive galaxies lie at the intersection of multiple filaments, but because the sub-halos lie along more than one filament, the projected distribution is typically not greatly elongated.

We therefore conclude that Lyman-$\alpha$ blobs are not in general strongly elongated (though clearly a few are). Observed samples are biased towards viewing elongated blobs down their long-axis, but this bias is not sufficient to explain the observed predominance of nearly circular blobs.

\section{Conclusions}

As our observations of Lyman-$\alpha$ blobs become better, the blobs appear more complicated. There is already evidence for multiple energy sources in different blobs \citep[e.g.][]{Colbert:2p629}. This paper suggests that for one blob at least, the origin of the surrounding gas is also complicated, combining merger-driven dense molecular gas in the nuclei, hot gas perhaps from a superwind and infalling lumpy colder gas. Our observations are consistent with seeing a massive galaxy being assembled by minor mergers, as described by \citet{2005MNRAS.363....2K}, \citet{Johansson:2009bb} and \cite{Oser:2011bp}, though with a major merger taking place at the
same time.

Cosmological hydrodynamical models which include some combination of winds and infall are rapidly increasing in sophistication \citep[e.g.][]{Zheng:2010p623,Goerdt:2010p611,Schaerer:2011p645,FaucherGiguere:2010p608,Zheng:2011p609} but most fail to predict the narrowness of the velocity sub-structures we see. The exception are the models of \citet{Goerdt:2010p611}, but the relative narrowness of the emission they predict is likely an artefact of the lack of radiative transfer in their models.

We conclude that high spectral resolution observations are providing an interesting challenge to the models. Few instruments are set up to make these high resolution observations in the blue, but those that can should pursue this.

\section*{Acknowledgments}

We'd like to thank Ralph Sutherland for several useful conversations.

The Millennium Simulation databases used in this paper and the web application providing online access to them were constructed as part of the activities of the German Astrophysical Virtual Observatory.

Based in part on observations obtained as part of program GS-2005B-Q-23 at the Gemini Observatory, which is operated by the Association of Universities for Research in Astronomy, Inc., under a cooperative agreement with the NSF on behalf of the Gemini partnership: the National Science Foundation (United States), the Science and Technology Facilities Council (United Kingdom), the National Research Council (Canada), CONICYT (Chile), the Australian Research Council (Australia), Minist\'{e}rio da Ci\^{\i}ncia, Tecnologia e Inova\c{c}\~{a}o (Brazil) band Ministerio de Ciencia, Tecnología e Innovaci\'{o}n Productiva (Argentina).

Based in part on observations made with the NASA/ESA Hubble Space Telescope, and obtained from the Hubble Legacy Archive, which is a collaboration between the Space Telescope Science Institute (STScI/NASA), the Space Telescope European Coordinating Facility (ST-ECF/ESA) and the Canadian Astronomy Data Centre (CADC/NRC/CSA).

Dopita acknowledges ARC support under Discovery project DP0984657.

\appendix

\section{Slit mask spectroscopy of other candidate blobs in the Field }

In this appendix, spectroscopy is presented of three other candidate Lyman-$\alpha$ emitters
which lie in the same supercluster as B1. All lie several arcminutes from B1 and so do
not directly impact on its physics. Their properties are, however, used in the various statistical arguments in the main body of the paper.

We had previously obtained spectra of most bright potential Lyman-$\alpha$ emitters in the supercluster surrounding B1 \citep{Francis:2004p634}, but three needed further observations. In this section we present spectroscopy of these three.

One, B5 in the list of \citet{Palunas:2004p630} was classified as a potential Lyman-$\alpha$ blob, albeit an unusual one. It consisted of a luminous narrow-band emitting core surrounded by faint narrow-band excess fuzz. Our previous spectrum of this object was ambiguous \citep{Francis:2004p634}: we saw a potential Lyman-$\alpha$ line but also two other possible weak lines which did not match. The second object, identified by \citet{Palunas:2004p630} and called B29 was the most luminous compact potential Lyman-$\alpha$ emitter in the field, but had not been previously observed due to fibre collision constraints with the spectrograph. It is located at 21:43:05.90 $-$44:27:21.0 (J2000). The third, B30, was a fainter narrow-band excess object found by Palunas et al, also not observed previously due to fibre placement issues. It is located at 21:43:06.42 $-$44:27:00.6 (J2000).

B5, B29 and B30 were observed with the GMOS spectrograph on the Gemini South Telescope. The observational set-up with GMOS is described by \citet{Scarlata:2009p340}: both objects were observed simultaneously for an exposure time of 8,800 seconds, in 0.6 arcsecond seeing, on the night of 2005 October 9th.

From B5, we detect a single very strong emission line at 410.0 nm, with a velocity width of $1000 {\rm km\ s}^{-1}$. The tentative additional lines seen by \citet{Francis:2004p634} were ruled out. On the basis of the line width, extended emission and equivalent width, so the line must be Lyman-$\alpha$. The Lyman-$\alpha$ emission was slightly extended along the slit, so we confirm that B5 is an extremely luminous extended Lyman-$\alpha$ emitting object at $z=2.38$.

B29 turned out to be a QSO at the supercluster redshift. It showed strong broad Lyman-$\alpha$ peaking at around 412.0 nm, as well as a number of other typical QSO broad lines. The line widths were around 440 nm (full width at half maximum height). Strong associated absorption was seen in Lyman-$\alpha$ and C~IV (154.9 nm), slightly red-shifted with respect to the emission-line peak.

B30 showed a single strong emission line at 412.2 nm, with a velocity width of $900{\rm km\ s}^{-1}$, and is hence confirmed as a Lyman-$\alpha$ emitting galaxy in the supercluster.

\bibliographystyle{mn2e}
\bibliography{blobpapers}

\begin{thebibliography}{61}
\expandafter\ifx\csname natexlab\endcsname\relax\def\natexlab#1{#1}\fi

\bibitem[{Adelberger {et~al}\mbox{.}(2006)Adelberger, Steidel, Kollmeier, \&
  Reddy}]{Adelberger:2006ii}
Adelberger K.~L., Steidel C.~C., Kollmeier J.~A., Reddy N.~A., 2006, The
  Astrophysical Journal, 637, 74

\bibitem[{Allen {et~al}\mbox{.}(2008)Allen, Groves, Dopita, Sutherland, \&
  Kewley}]{Allen:2008gj}
Allen M.~G., Groves B.~A., Dopita M.~A., Sutherland R.~S., Kewley L.~J., 2008,
  Monthly Notices of the Royal Astronomical Society, 178, 20

\bibitem[{Bower {et~al}\mbox{.}(2004)Bower, Morris, Bacon, Wilman, Sullivan,
  Chapman, Davies, de~Zeeuw, \& Emsellem}]{Bower:2004p577}
Bower R.~G. {et~al.}, 2004, Monthly Notices of the Royal Astronomical Society,
  351, 63

\bibitem[{Boylan-Kolchin {et~al}\mbox{.}(2009)Boylan-Kolchin, Springel, White,
  Jenkins, \& Lemson}]{BoylanKolchin:2009co}
Boylan-Kolchin M., Springel V., White S. D.~M., Jenkins A., Lemson G., 2009,
  Monthly Notices of the Royal Astronomical Society, 398, 1150

\bibitem[{Bruzual \& Charlot(2003)}]{Bruzual:2003ck}
Bruzual G., Charlot S., 2003, Monthly Notices of the Royal Astronomical
  Society, 344, 1000

\bibitem[{Calzetti {et~al}\mbox{.}(1994)Calzetti, Kinney, \&
  Storchi-Bergmann}]{1994ApJ...429..582C}
Calzetti D., Kinney A.~L., Storchi-Bergmann T., 1994, The Astrophysical
  Journal, 429, 582

\bibitem[{Chapman {et~al}\mbox{.}(2004)Chapman, Scott, Windhorst, Frayer,
  Borys, Lewis, \& Ivison}]{Chapman:2004fq}
Chapman S.~C., Scott D., Windhorst R.~A., Frayer D.~T., Borys C., Lewis G.~F.,
  Ivison R.~J., 2004, The Astrophysical Journal, 606, 85

\bibitem[{Christensen {et~al}\mbox{.}(2006)Christensen, Jahnke, Wisotzki, \&
  S{\'a}nchez}]{Christensen:2006p637}
Christensen L., Jahnke K., Wisotzki L., S{\'a}nchez S.~F., 2006, Astronomy and
  Astrophysics, 459, 717

\bibitem[{Colbert {et~al}\mbox{.}(2011)Colbert, Scarlata, Teplitz, Francis,
  Palunas, Williger, \& Woodgate}]{Colbert:2p629}
Colbert J., Scarlata C., Teplitz H., Francis P., Palunas P., Williger G.,
  Woodgate B., 2011, The Astrophysical Journal, 728, 59

\bibitem[{Colbert {et~al}\mbox{.}(2006)Colbert, Teplitz, Francis, Palunas,
  Williger, \& Woodgate}]{Colbert:2006p628}
Colbert J.~W., Teplitz H., Francis P., Palunas P., Williger G.~M., Woodgate B.,
  2006, The Astrophysical Journal, 637, L89

\bibitem[{De~Lucia \& Blaizot(2007)}]{DeLucia:2007ju}
De~Lucia G., Blaizot J., 2007, Monthly Notices of the Royal Astronomical
  Society, 375, 2

\bibitem[{Dijkstra {et~al}\mbox{.}(2006)Dijkstra, Haiman, \&
  Spaans}]{Dijkstra:2006dh}
Dijkstra M., Haiman Z., Spaans M., 2006, The Astrophysical Journal, 649, 14

\bibitem[{Dijkstra \& Loeb(2008)}]{Dijkstra:2008kd}
Dijkstra M., Loeb A., 2008, Monthly Notices of the Royal Astronomical Society,
  386, 492

\bibitem[{Dijkstra \& Loeb(2009)}]{Dijkstra:2009p602}
Dijkstra M., Loeb A., 2009, Monthly Notices of the Royal Astronomical Society,
  400, 1109

\bibitem[{Dopita {et~al}\mbox{.}(2007)Dopita, Hart, McGregor, Oates, Bloxham,
  \& Jones}]{2007Ap&SS.310..255D}
Dopita M., Hart J., McGregor P., Oates P., Bloxham G., Jones D., 2007,
  Astrophysics and Space Science, 310, 255

\bibitem[{Dopita {et~al}\mbox{.}(2010)Dopita, Rhee, Farage, McGregor, Bloxham,
  Green, Roberts, Neilson, Wilson, Young, Firth, Busarello, \&
  Merluzzi}]{Dopita:2010fc}
Dopita M. {et~al.}, 2010, Astrophysics and Space Science, 327, 245

\bibitem[{Erb {et~al}\mbox{.}(2011)Erb, Bogosavljevi{\'c}, \&
  Steidel}]{Erb:2011p612}
Erb D.~K., Bogosavljevi{\'c} M., Steidel C.~C., 2011, The Astrophysical Journal
  Letters, 740, L31

\bibitem[{Faucher-Gigu{\`e}re {et~al}\mbox{.}(2010)Faucher-Gigu{\`e}re, Kere{\v
  s}, Dijkstra, Hernquist, \& Zaldarriaga}]{FaucherGiguere:2010p608}
Faucher-Gigu{\`e}re C.-A., Kere{\v s} D., Dijkstra M., Hernquist L.,
  Zaldarriaga M., 2010, The Astrophysical Journal, 725, 633

\bibitem[{Francis \& Bland-Hawthorn(2004)}]{Francis:2004dqa}
Francis P.~J., Bland-Hawthorn J., 2004, Monthly Notices of the Royal
  Astronomical Society, 353, 301

\bibitem[{Francis \& Hewett(1993)}]{1993AJ....105.1633F}
Francis P.~J., Hewett P.~C., 1993, Astronomical Journal (ISSN 0004-6256), 105,
  1633

\bibitem[{Francis \& McDonnell(2006)}]{Francis:2006cya}
Francis P.~J., McDonnell S., 2006, Monthly Notices of the Royal Astronomical
  Society, 370, 1372

\bibitem[{Francis {et~al}\mbox{.}(2004{\natexlab{a}})Francis, Nelson, \&
  Cutri}]{Francis:2004ev}
Francis P.~J., Nelson B.~O., Cutri R.~M., 2004{\natexlab{a}}, The Astronomical
  Journal, 127, 646

\bibitem[{Francis {et~al}\mbox{.}(2004{\natexlab{b}})Francis, Palunas, Teplitz,
  Williger, \& Woodgate}]{Francis:2004p634}
Francis P.~J., Palunas P., Teplitz H.~I., Williger G.~M., Woodgate B.~E.,
  2004{\natexlab{b}}, The Astrophysical Journal, 614, 75

\bibitem[{Francis \& Williger(2004)}]{Francis:2004jg}
Francis P.~J., Williger G.~M., 2004, The Astrophysical Journal, 602, L77

\bibitem[{Francis {et~al}\mbox{.}(2001)Francis, Williger, Collins, Palunas,
  Malumuth, Woodgate, Teplitz, Smette, Sutherland, Danks, Hill, Lindler,
  Kimble, Heap, \& Hutchings}]{Francis:2001jb}
Francis P.~J. {et~al.}, 2001, The Astrophysical Journal, 554, 1001

\bibitem[{Francis {et~al}\mbox{.}(1997)Francis, Woodgate, \&
  Danks}]{1997ApJ...482L..25F}
Francis P.~J., Woodgate B.~E., Danks A.~C., 1997, The Astrophysical Journal,
  482, L25

\bibitem[{Francis {et~al}\mbox{.}(1996)Francis, Woodgate, Warren, Moller,
  Mazzolini, Bunker, Lowenthal, Williams, Minezaki, Kobayashi, \&
  Yoshii}]{Francis:1996p627}
Francis P.~J. {et~al.}, 1996, Astrophysical Journal v.457, 457, 490

\bibitem[{Goerdt {et~al}\mbox{.}(2010)Goerdt, Dekel, Sternberg, Ceverino,
  Teyssier, \& Primack}]{Goerdt:2010p611}
Goerdt T., Dekel A., Sternberg A., Ceverino D., Teyssier R., Primack J.~R.,
  2010, Monthly Notices of the Royal Astronomical Society, 407, 613

\bibitem[{Guaita {et~al}\mbox{.}(2010)Guaita, Gawiser, Padilla, Francke, Bond,
  Gronwall, Ciardullo, Feldmeier, Sinawa, Blanc, \& Virani}]{Guaita:2010kr}
Guaita L. {et~al.}, 2010, The Astrophysical Journal, 714, 255

\bibitem[{Haiman \& Rees(2001)}]{Haiman:2001bb}
Haiman Z., Rees M.~J., 2001, The Astrophysical Journal, 556, 87

\bibitem[{Hayes {et~al}\mbox{.}(2011)Hayes, Scarlata, \&
  Siana}]{Hayes:2011p606}
Hayes M., Scarlata C., Siana B., 2011, Nature, 476, 304

\bibitem[{Hinshaw {et~al}\mbox{.}(2009)Hinshaw, Weiland, Hill, Odegard, Larson,
  Bennett, Dunkley, Gold, Greason, Jarosik, Komatsu, Nolta, Page, Spergel,
  Wollack, Halpern, Kogut, Limon, Meyer, Tucker, \& Wright}]{Hinshaw:2009jq}
Hinshaw G. {et~al.}, 2009, Monthly Notices of the Royal Astronomical Society,
  180, 225

\bibitem[{Johansson {et~al}\mbox{.}(2009)Johansson, Naab, \&
  Ostriker}]{Johansson:2009bb}
Johansson P.~H., Naab T., Ostriker J.~P., 2009, The Astrophysical Journal, 697,
  L38

\bibitem[{Kennicutt~Jr.(1998{\natexlab{a}})}]{KennicuttJr:1998ki}
Kennicutt~Jr. R.~C., 1998{\natexlab{a}}, Annual Review of Astronomy and
  Astrophysics, 36, 189

\bibitem[{Kennicutt~Jr.(1998{\natexlab{b}})}]{KennicuttJr:1998id}
Kennicutt~Jr. R.~C., 1998{\natexlab{b}}, The Astrophysical Journal, 498, 541

\bibitem[{Kere{\v s} {et~al}\mbox{.}(2005)Kere{\v s}, Katz, Weinberg, \&
  Dav{\'e}}]{2005MNRAS.363....2K}
Kere{\v s} D., Katz N., Weinberg D.~H., Dav{\'e} R., 2005, Monthly Notices of
  the Royal Astronomical Society, 363, 2

\bibitem[{King {et~al}\mbox{.}(2011)King, Zubovas, \& Power}]{King:2011ey}
King A.~R., Zubovas K., Power C., 2011, Monthly Notices of the Royal
  Astronomical Society: Letters, 415, L6

\bibitem[{Matsuda {et~al}\mbox{.}(2004)Matsuda, Yamada, Hayashino, Tamura,
  Yamauchi, Ajiki, Fujita, Murayama, Nagao, Ohta, Okamura, Ouchi, Shimasaku,
  Shioya, \& Taniguchi}]{Matsuda:2004p631}
Matsuda Y. {et~al.}, 2004, The Astronomical Journal, 128, 569

\bibitem[{Oser {et~al}\mbox{.}(2011)Oser, Naab, Ostriker, \&
  Johansson}]{Oser:2011bp}
Oser L., Naab T., Ostriker J.~P., Johansson P.~H., 2011, The Astrophysical
  Journal, 744, 63

\bibitem[{Palunas {et~al}\mbox{.}(2004)Palunas, Teplitz, Francis, Williger, \&
  Woodgate}]{Palunas:2004p630}
Palunas P., Teplitz H.~I., Francis P.~J., Williger G.~M., Woodgate B.~E., 2004,
  The Astrophysical Journal, 602, 545

\bibitem[{Prescott {et~al}\mbox{.}(2011{\natexlab{a}})Prescott, Dey, Brodwin,
  Chaffee, Desai, Eisenhardt, Le~Floc'h, Jannuzi, Kashikawa, Matsuda, \&
  Soifer}]{Prescott:2011wl}
Prescott M. K.~M. {et~al.}, 2011{\natexlab{a}}, eprint arXiv:1111.0630

\bibitem[{Prescott {et~al}\mbox{.}(2011{\natexlab{b}})Prescott, Smith, Schmidt,
  \& Dey}]{Prescott:2011p607}
Prescott M. K.~M., Smith P.~S., Schmidt G.~D., Dey A., 2011{\natexlab{b}}, The
  Astrophysical Journal Letters, 730, L25

\bibitem[{Reuland {et~al}\mbox{.}(2007)Reuland, van Breugel, de~Vries, Dopita,
  Dey, Miley, R{\"o}ttgering, Venemans, Stanford, Lacy, Spinrad, Dawson, Stern,
  \& Bunker}]{Reuland:2007et}
Reuland M. {et~al.}, 2007, The Astronomical Journal, 133, 2607

\bibitem[{Reuland {et~al}\mbox{.}(2003)Reuland, van Breugel, R{\"o}ttgering,
  de~Vries, Stanford, Dey, Lacy, Bland-Hawthorn, Dopita, \&
  Miley}]{Reuland:2003gz}
Reuland M. {et~al.}, 2003, The Astrophysical Journal, 592, 755

\bibitem[{Scarlata {et~al}\mbox{.}(2009)Scarlata, Colbert, Teplitz, Bridge,
  Francis, Palunas, Siana, Williger, \& Woodgate}]{Scarlata:2009p340}
Scarlata C. {et~al.}, 2009, The Astrophysical Journal, 706, 1241

\bibitem[{Schaerer {et~al}\mbox{.}(2011)Schaerer, Hayes, Verhamme, \&
  Teyssier}]{Schaerer:2011p645}
Schaerer D., Hayes M., Verhamme A., Teyssier R., 2011, Astronomy and
  Astrophysics, 531, 12

\bibitem[{Springel {et~al}\mbox{.}(2005)Springel, White, Jenkins, Frenk,
  Yoshida, Gao, Navarro, Thacker, Croton, Helly, Peacock, Cole, Thomas,
  Couchman, Evrard, Colberg, \& Pearce}]{Springel:2005gv}
Springel V. {et~al.}, 2005, Nature, 435, 629

\bibitem[{Steidel {et~al}\mbox{.}(2000)Steidel, Adelberger, Shapley, Pettini,
  Dickinson, \& Giavalisco}]{Steidel:2000bw}
Steidel C.~C., Adelberger K.~L., Shapley A.~E., Pettini M., Dickinson M.,
  Giavalisco M., 2000, The Astrophysical Journal, 532, 170

\bibitem[{Steidel {et~al}\mbox{.}(2011)Steidel, Bogosavljevi{\'c}, Shapley,
  Kollmeier, Reddy, Erb, \& Pettini}]{Steidel:2011jk}
Steidel C.~C., Bogosavljevi{\'c} M., Shapley A.~E., Kollmeier J.~A., Reddy
  N.~A., Erb D.~K., Pettini M., 2011, The Astrophysical Journal, 736, 160

\bibitem[{Taniguchi {et~al}\mbox{.}(2001)Taniguchi, Shioya, \&
  Kakazu}]{Taniguchi:2001fq}
Taniguchi Y., Shioya Y., Kakazu Y., 2001, The Astrophysical Journal, 562, L15

\bibitem[{Treister {et~al}\mbox{.}(2009)Treister, Urry, \&
  Virani}]{Treister:2009p653}
Treister E., Urry C.~M., Virani S., 2009, The Astrophysical Journal, 696, 110

\bibitem[{Veilleux {et~al}\mbox{.}(2005)Veilleux, Cecil, \&
  Bland-Hawthorn}]{Veilleux:2005ec}
Veilleux S., Cecil G., Bland-Hawthorn J., 2005, Annual Review of Astronomy and
  Astrophysics, 43, 769

\bibitem[{Webster {et~al}\mbox{.}(1995)Webster, Francis, Petersont, Drinkwater,
  \& Masci}]{Webster:1995fb}
Webster R.~L., Francis P.~J., Petersont B.~A., Drinkwater M.~J., Masci F.~J.,
  1995, Nature, 375, 469

\bibitem[{Weidinger {et~al}\mbox{.}(2005)Weidinger, M{\o}ller, Fynbo, \&
  Thomsen}]{Weidinger:2005p636}
Weidinger M., M{\o}ller P., Fynbo J. P.~U., Thomsen B., 2005, Astronomy and
  Astrophysics, 436, 825

\bibitem[{Weijmans {et~al}\mbox{.}(2010)Weijmans, Bower, Geach, Swinbank,
  Wilman, de~Zeeuw, \& Morris}]{Weijmans:2010p583}
Weijmans A.-M., Bower R.~G., Geach J.~E., Swinbank A.~M., Wilman R.~J.,
  de~Zeeuw P.~T., Morris S.~L., 2010, Monthly Notices of the Royal Astronomical
  Society, 402, 2245

\bibitem[{Wilman {et~al}\mbox{.}(2005)Wilman, Gerssen, Bower, Morris, Bacon,
  de~Zeeuw, \& Davies}]{Wilman:2005p582}
Wilman R.~J., Gerssen J., Bower R.~G., Morris S.~L., Bacon R., de~Zeeuw P.~T.,
  Davies R.~L., 2005, Nature, 436, 227

\bibitem[{Yang {et~al}\mbox{.}(2011)Yang, Zabludoff, Jahnke, Eisenstein,
  Dav{\'e}, Shectman, \& Kelson}]{Yang:2011hz}
Yang Y., Zabludoff A., Jahnke K., Eisenstein D., Dav{\'e} R., Shectman S.~A.,
  Kelson D.~D., 2011, The Astrophysical Journal, 735, 87

\bibitem[{Yang {et~al}\mbox{.}(2009)Yang, Zabludoff, Tremonti, Eisenstein, \&
  Dav{\'e}}]{Yang:2009p635}
Yang Y., Zabludoff A., Tremonti C., Eisenstein D., Dav{\'e} R., 2009, The
  Astrophysical Journal, 693, 1579

\bibitem[{Zheng {et~al}\mbox{.}(2010)Zheng, Cen, Trac, \&
  Miralda-Escud{\'e}}]{Zheng:2010p623}
Zheng Z., Cen R., Trac H., Miralda-Escud{\'e} J., 2010, The Astrophysical
  Journal, 716, 574

\bibitem[{Zheng {et~al}\mbox{.}(2011)Zheng, Cen, Weinberg, Trac, \&
  Miralda-Escud{\'e}}]{Zheng:2011p609}
Zheng Z., Cen R., Weinberg D., Trac H., Miralda-Escud{\'e} J., 2011, The
  Astrophysical Journal, 739, 62

\bibitem[{Zheng \& Miralda-Escud{\'e}(2002)}]{Zheng:2002p624}
Zheng Z., Miralda-Escud{\'e} J., 2002, The Astrophysical Journal, 578, 33

\end{thebibliography}

\label{lastpage}

\end{document}